\documentclass[twocolumn,aps,prx,floatfix]{revtex4-2}
\pdfoutput=1
\usepackage[T1]{fontenc}
\usepackage{amsmath,amssymb,amscd}
\usepackage{graphicx}
\usepackage[leftcaption]{sidecap}
\usepackage{tikz}
\usetikzlibrary{arrows.meta,decorations.pathreplacing,decorations.pathmorphing}
\usepackage[colorlinks,allcolors=blue]{hyperref}

\usepackage[most]{tcolorbox}
\newtcbox{\othermathbox}[1][]{nobeforeafter, math upper, tcbox raise base, enhanced, rounded corners, colback=black!5, colframe=black, left=0.7em, top=0.4em, right=0.7em, bottom=0.5em}
\usepackage{empheq}

\usepackage{color}
\usepackage{enumitem}
\usepackage{stmaryrd}
\usepackage{mathrsfs}
\usepackage{amsfonts}
\usepackage{amssymb,scalerel,stackengine}
\newcommand\beq{\begin{equation}}
\newcommand\ee{\end{equation}}
\newcommand\cO{{\cal O}}

\newcommand\dd{\text{d}}

\newcommand\ie{{\it i.e.}\ }

\def\hi{{\hat i}}

\def\ha{{\hat a}}
\def\hb{{\hat b}}

\def\hr{{\hat r}}

\def\pd{\partial}

\def\eps{\epsilon}

\def\cJ{\mathcal{J}}
\def\cM{\mathcal{M}}

\def\cF{\mathcal{F}}

\def\cO{\mathcal{O}}

\def\bS{\boldsymbol{S}}

\def\th{\theta}

\def\tl{\widetilde}

\newcommand\be[1]{{\boldsymbol{e}_{\hat{#1}}}}
\newcommand\bff[1]{{\boldsymbol{f}_{\hat{#1}}}}

\newcommand\E[2]{{\zeta_{\hat{#1}}^{\;\;#2}}}

\def\bu{{\boldsymbol u}}

\newcommand{\letter}{letter}

\begin{document}

\title{The Precession Caused by Gravitational Waves}

%\author{Authors}

\author{Ali Seraj}
\affiliation{Centre for Gravitational Waves, Universit\'e Libre de Bruxelles; International Solvay Institutes, CP 231, B-1050 Brussels, Belgium}
\author{Blagoje Oblak}
\affiliation{CPHT, CNRS, Ecole Polytechnique, IP Paris, F-91128 Palaiseau;\\ LPTHE, CNRS, Sorbonne Universit\'e, 75252 Paris Cedex 05, France}

\begin{abstract}
We show that gravitational waves cause freely falling gyroscopes to precess relative to fixed distant stars, extending the stationary Lense-Thirring effect. The precession rate decays as the square of the inverse distance to the source, and is proportional to a suitable Noether current for dual asymptotic symmetries at null infinity. Integrating the rate over time yields a net rotation---a `gyroscopic memory'---whose angle reproduces the known spin memory effect but also contains an extra contribution due to the generator of gravitational electric-magnetic duality. The angle's order of magnitude for the first LIGO signal is estimated to be $\Phi\sim 10^{-35}$ arcseconds near Earth, but the effect may be substantially larger for supermassive black hole mergers.
\end{abstract}

\maketitle

\section{Introduction}
\label{sintro}

\begin{figure}[b]
\centering
\includegraphics[width=.5\columnwidth]{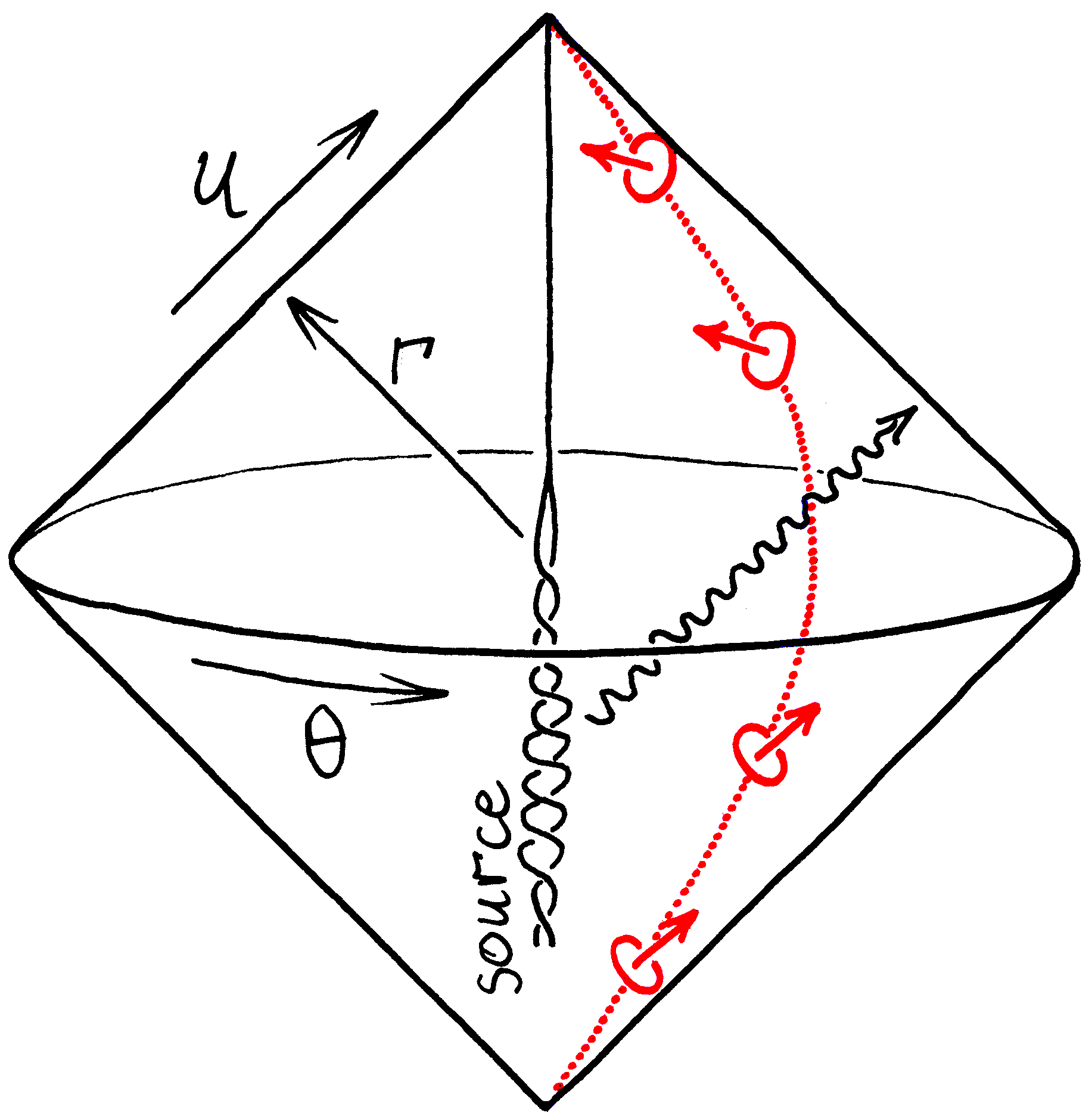}
\caption{The world-line of a freely falling observer with a gyroscope, represented here (in red) in a Penrose diagram of near-Minkowski space. The gyroscope's spin is parallel-transported along its trajectory, but the passage of gravitational waves causes its orientation to change relative to fixed distant stars. Bondi coordinates $(u,r,\th^a)$ are included; the source of radiation is located at the origin $r=0$.}
\label{fiZZ}
\end{figure}

Consider an observer floating freely in outer space, carrying a spinning gyroscope. The observer looks at fixed distant stars in order to measure the gyroscope's orientation. When a localized source emits a burst of gravitational waves that crosses the observer's path, the gyroscope precesses and eventually settles in a new orientation (Fig.\ \ref{fiZZ}); the corresponding rotation angle carries a `memory' of the wave profile. The goal of this \letter\ is to describe this precession and the ensuing memory effect.

\begin{figure}[t]
\centering
\includegraphics[width=\columnwidth]{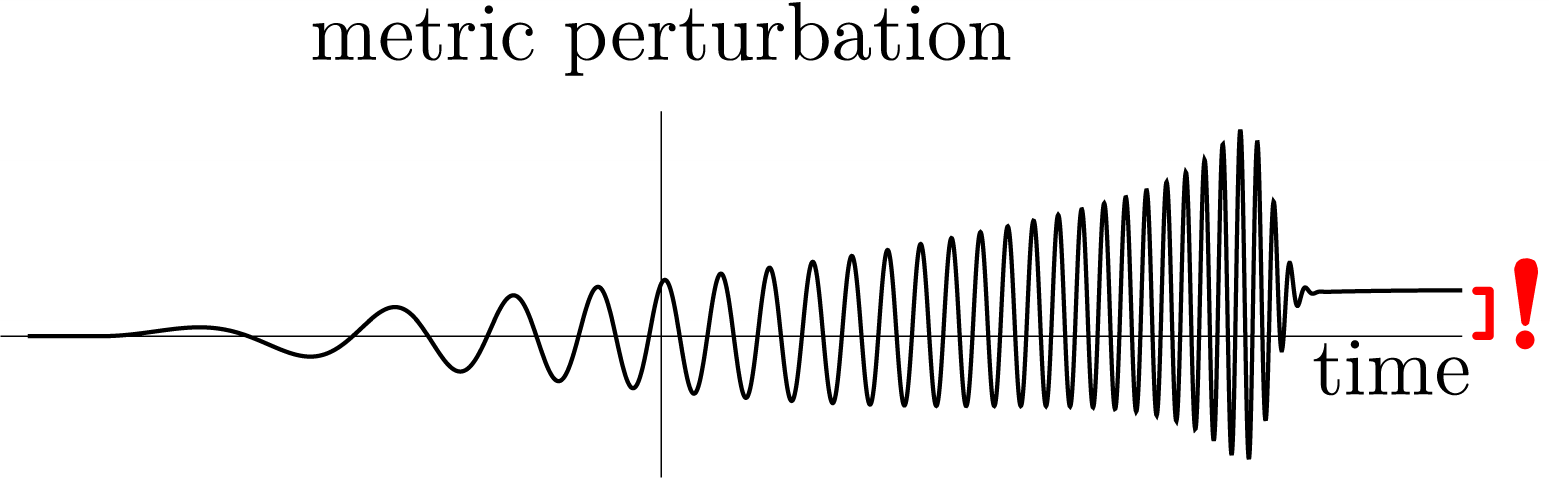}
\caption{A cartoon of the typical local metric perturbation caused by gravitational waves. Even after the end of the disturbance, some metric components (typically some function of the shear $C_{ab}$ in Eq.\ \eqref{s6}) differ from their initial value by an amount that depends on the waveform. This net offset has potentially observable consequences; one of them is the gyroscopic memory described here. See section \ref{se3b} for a more detailed discussion of this plot in the gyroscopic context.}
\label{fiMem}
\end{figure}

The motivations for this investigation are twofold. The first is the recent breakthrough observation of gravitational waves \cite{LIGOScientific:2016aoc}, which makes it realistic to seek their measurable signatures. In particular, {\it memory effects} \cite{Zeldovich:1974gvh,Braginsky:1985vlg,braginsky1987gravitational,Christodoulou:1991cr,Blanchet:1992br,Zhang:2017rno,Zhang:2017geq,Flanagan:2018yzh,Divakarla:2021xrd,Pasterski:2015tva,Mao:2018xcw} sensitive to the net offset of metric components after a gravitational wave burst (see Fig.\ \ref{fiMem}) may be observable in the near future \cite{Favata:2010zu,Lasky:2016knh,Boersma:2020gxx}. The second motivation has to do with fundamental symmetries of classical and quantum gravity. Indeed, asymptotically Minkowskian space-time metrics enjoy an infinite-dimensional `Bondi-Metzner-Sachs symmetry' \cite{Bondi:1962px,Sachs2,Barnich:2009se,Barnich:2010eb} whose Noether currents at null infinity were recently related to the displacement memory that affects nearby freely falling test masses \cite{Strominger:2014pwa}. In a similar vein, the rate of gyroscopic precession found here turns out to coincide with a current \cite{Freidel:2021fxf,Freidel:2021qpz} for so-called `dual supertranslations' \cite{Godazgar:2018qpq,Godazgar:2018dvh,Kol:2019nkc,Godazgar:2019dkh,Godazgar:2020gqd,Godazgar:2020kqd,Oliveri:2020xls,Kol:2020ucd}. Furthermore, the net change of orientation before and after the passage of waves involves a `superrotation' charge \cite{Campiglia:2014yka,Compere:2018ylh,Compere:2019gft} and a generator of local gravitational electric-magnetic duality transformations. To our knowledge, this is the first appearance of such dual symmetries in a simple local measurement protocol for gravitational waves.

There is in fact a third, perhaps more academic, motivation for this work. Indeed, while our results are related to deep properties of the gravitational phase space at the forefront of research, our method is comparatively elementary: it consists in rewriting the parallel transport equation of a spin vector in a radiative space-time, with respect to a tetrad whose elements point towards fixed distant stars. The computation is thus an `exercise' that generalizes Lense-Thirring precession \cite[sec.\ 40.7]{Misner:1973prb} to radiative metrics, and it could have been carried out sixty years ago \cite{Bondi:1962px,Sachs2}. It seems important indeed to fill such a gap in the literature.

The \letter\ is organized as follows. First, section \ref{se2} sets the stage by displaying the (Bondi) coordinates and metric \cite{Bondi:1962px,Sachs2} to be used throughout, and contains a description of the tetrad with respect to which precession is to be evaluated. Section \ref{se3} is then devoted to our results, namely the expression of the precession rate in terms of a `dual covariant mass aspect' \cite{Freidel:2021fxf,Freidel:2021qpz} and that of orientation memory in terms of a surface charge, its flux, and a generator of electric-magnetic duality. Aside from the latter, this actually reproduces the spin memory effect \cite{Pasterski:2015tva} as a special case. Note that computational details are omitted throughout: they are relegated to the longer companion paper \cite{Seraj:2021rxd}.

\section{Metric and tetrad}
\label{se2}

This section introduces Bondi coordinates, the corresponding asymptotically flat metrics, and the tetrad that will be used in section \ref{se3} to measure the gyroscope's orientation relative to fixed distant stars. Readers familiar with the Bondi metric ansatz may jump straight to the discussion of the tetrad in section \ref{se2b}.

\subsection{Bondi coordinates and metric}
\label{se2a}

Choose an origin in space and label the points of the four-dimensional space-time manifold by {\it retarded Bondi coordinates}: an areal distance $r$, a retarded time $u$, and angular coordinates $\th^a$ ($a=1,2$) on a unit celestial sphere with metric $q_{ab}(\th)\,\dd\th^a\dd\th^b$ (see Fig.\ \ref{fiZZ}). Any outgoing light-like ray then propagates so as to eventually reach {\it future null infinity}, \ie the region $r\to\infty$ with finite $u$. Accordingly, we assume throughout that our observer is located at large $r$. It is then meaningful to expand the components of the space-time metric as asymptotic series in $1/r$ \cite{Flanagan:2015pxa}:
\beq
\label{s6}
\begin{split}
\dd s^2
\sim&
-\big(1-\tfrac{2m}{r}-\tfrac{2F}{r^2}\big)\dd u^2-2\big(1-\tfrac{C^2}{16r^2}\big)\dd u\,\dd r\\
&+\big(r^2q_{ab}+rC_{ab}+\tfrac{1}{4}q_{ab}C^2\big)\dd\th^a\,\dd\th^b\\
&+2\big(\tfrac{1}{2}D^bC_{ab}+\tfrac{1}{r}\left[\tfrac{2}{3}L_a-\tfrac{1}{16}\pd_a(C^2)\right]
\big)\dd u\,\dd\th^a,
\end{split}
\ee
where the functions $m$, $F$, $C_{ab}$ and $L_a$ only depend on $(u,\th)$. (We also write $C^2\equiv C_{ab}C^{ab}$ to reduce clutter, with indices raised and lowered thanks to the metric $q_{ab}$; $D_a$ is the covariant derivative on $S^2$.) The radial dependence is thus explicit and the metric reduces to the pure Minkowski form $\dd s^2=-\dd u^2-2\,\dd u\,\dd r+r^2q_{ab}\,\dd\th^a\,\dd\th^b$ in the limit $r\to\infty$.

The most important quantity in \eqref{s6} is the (symmetric and traceless) {\it asymptotic shear} tensor $C_{ab}(u,\th)$, which contains all the information about radiation. Its time-dependence is unconstrained and determines the {\it news tensor} $N_{ab}\equiv\partial_uC_{ab}$ that will play a key role below. The function $m(u,\th)$ is the {\it Bondi mass aspect} and $L_a(u,\th)$ is the {\it angular momentum aspect}, respectively measuring `densities' of energy and angular momentum at null infinity. Their time-dependence is fixed by the shear and news tensors through so-called {\it balance equations} \cite{Flanagan:2015pxa}
\begin{subequations}
\label{t6}
\begin{align}
\dot{m}
&=
\tfrac{1}{4}D_aD_bN^{ab}-\tfrac{1}{8} N_{ab}N^{ab},\\
\label{tt6}
\dot{L}_{a}
&=
D_am+\tfrac{1}{2}D^bD_{[a}D^cC_{b]c}
-\cJ_a,
\end{align}
\end{subequations}
where $\dot X\equiv\partial_uX$ while $\cJ$ is a local quadratic flux
\beq
\label{fluxes}
%\cP
%\equiv
%\tfrac{1}{8} N_{ab}N^{ab},
%\quad
\cJ_a
\equiv
{-}\tfrac{1}{4} D_b(N^{bc} C_{ac}){-}\tfrac{1}{2} D_bN^{bc}C_{ac}
\ee
that will turn out to affect orientation memory in section \ref{se3b}. The remaining function $F$ in \eqref{s6} is then given by $F=-\tfrac{1}{32}C^2-\tfrac{1}{6}\left(D_{a} L^{a}\right)-\tfrac{1}{8}(D\cdot C)^2$, where we let $(D\cdot C)^2\equiv D_bC^{ab}D^cC_{ac}$ for brevity.

The full metric \eqref{s6} contains numerous subleading corrections in $1/r$, all of which we omit since they will play no role below. Crucially, all subleading terms are determined by leading metric data up to time-independent `integration functions' on celestial spheres \cite{Barnich:2010eb,Grant:2021hga}. This is similar to mass and angular momentum, whose time evolution \eqref{t6} is entirely fixed by news so that only the initial conditions $m(u_0,\th)$ and $L_a(u_0,\th)$ are arbitrary.

\subsection{Tetrad carried by a freely falling observer}
\label{se2b}

Consider a freely falling observer at large $r$ in an asymptotically flat metric \eqref{s6}. We wish to build an orthonormal tetrad $\{\bff{\mu}|\mu=0,1,2,3\}$ such that $\bff{0}=\bu$ is the observer's proper velocity while $\bff{i}$ ($i=1,2,3$) are space-like vectors pointing towards fixed `distant stars' at infinity. (Hatted indices label tetrad elements, and they are raised/lowered using the inertial Minkowski metric.) In practice, Bondi coordinates heavily rely on a choice of origin---the location of the source of radiation. Accordingly, we first build a `source-oriented' tetrad $\{\be{\mu}\}$, then perform angle-dependent rotations so as to produce the desired frame $\{\bff{\mu}\}$, which we call `star-oriented'.

Our starting point is the observer's proper velocity
\beq
\label{s7}
\be{0}
=
\bff{0}
=
\bu
=
\gamma(\pd_u+v^r\pd_r+v^a\pd_a),
\ee
\ie the zeroth element of both tetrads. In these terms, solving the geodesic equation for the metric \eqref{s6} with the condition $\bu\sim\partial_u$ at large $r$ (the observer is asymptotically at rest) yields
\begin{subequations}
\begin{align}
\label{geu}
\gamma
&=
1+\tfrac{m_0}{r}+\tfrac{\gamma_2}{r^2},
\quad
\gamma_2
\equiv
\tfrac{1}{16}C^2+\smallint \!m\\
%\int_{u_0}^u du'\,m,\\
\label{ger}
v^r
&=
\tfrac{m-m_0}{r}+\tfrac{1}{r^2}\big(-\gamma_2-\tfrac{1}{6}D_aL^a-\tfrac{1}{8}(D\cdot C)^2\big),\\
\label{gea}
v^a
&=
{-}\tfrac{1}{2r^2}D_bC^{ab}{+}\tfrac{1}{r^3}\big(D^a\gamma_2{-}\tfrac23L^a{+}\tfrac12C^{ab}D^cC_{bc}\big),
\end{align}
\end{subequations}
with $m_0\equiv m(u_0,\th)$ the initial Bondi mass aspect and $\int m\equiv \int_{u_0}^u du'\,m$. Only spatial tetrad elements thus remain to be found. In the source-oriented case, one first obtains the radial vector $\be{r}$ by following an outgoing null radial geodesic, projecting out the $\bu$ component of its velocity and finally expanding in $1/r$, which yields $\be{r}\sim\tfrac{1}{\gamma}(1+\tfrac{1}{r^2}C^2/16)\partial_r-\bu$. The tetrad is then completed by picking an orthonormal dyad $\E{a}{\!}$ on the unit sphere and expanding angular tetrad elements as
\beq
\label{s8}
\be{a}
\sim
\tfrac{1}{r}\E{a}{b}(\delta^a_b-\tfrac{1}{2r}C^a_b+\tfrac{1}{16r^2}C^2\delta^a_b)
(\partial_a+\tfrac{1}{r}D_a \smallint \!m\,\partial_r).
\ee
This furnishes an explicit source-oriented Lorentz frame $\{\be{0},\be{r},\be{a}\}$, written here perturbatively in $1/r$.

\begin{figure}[t]
\centering
\includegraphics[width=.8\columnwidth]{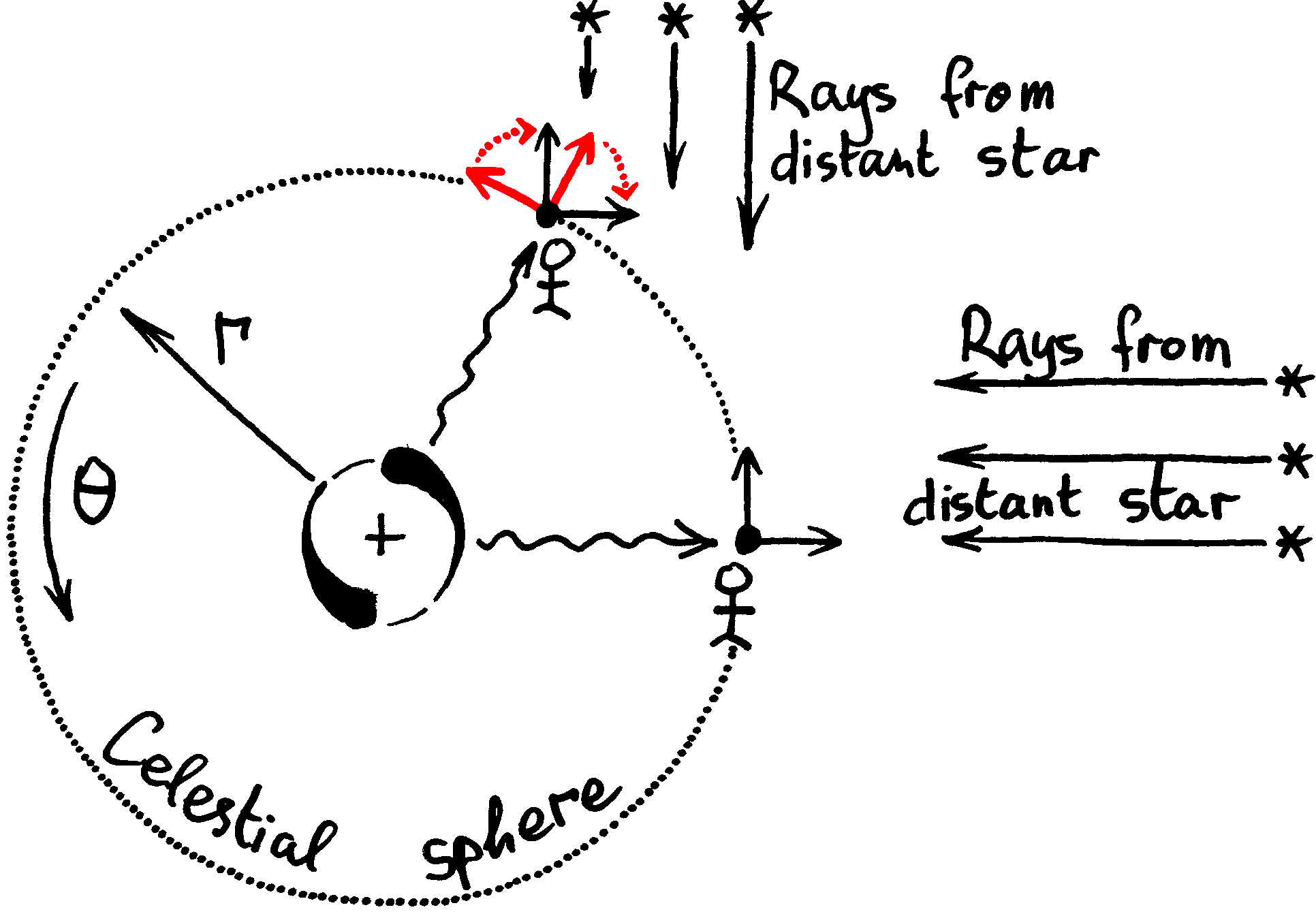}
\caption{In Bondi coordinates, the most natural tetrad $\{\be{\mu}\}$ is a source-oriented one (with a radial vector $\be{r}$ aligned with outgoing null geodesics). Converting such a frame into a tetrad $\{\bff{\mu}\}$ pointing towards fixed distant stars requires a local rotation $R(\th)$, as in Eq.\ \eqref{s9}.}
\label{fiXX}
\end{figure}

Obtaining a tetrad whose spatial vectors point towards fixed distant stars requires an extra step (see Fig.\ \ref{fiXX}), as the spatial vectors $\{\be{r},\be{a}\}$ need to be rotated in an angle-dependent manner. This is intuitively obvious: even in pure Minkowski space, the source-oriented tetrad of a freely falling observer must rotate continuously so that its vector $\be{r}$ points towards the origin. Any gyroscope with non-zero angular velocity trivially precesses relative to this tetrad, even without radiation. One may therefore compensate this effect by defining transformed tetrad vectors
\beq
\label{s9}
\bff{i}
=
R_{\hat i}{}^{\hat j}(\th)\,\be{j},
\ee
where the local rotation matrix is a path-ordered exponential $R(\th)=P\exp\int_{\th_0}^{\th}\overline{\omega}$ of the spin connection $\overline{\omega}$ of the flat space triad $\{\partial_r-\pd_u,\tfrac{1}{r}\E{a}{\!}\}$, with $\E{a}{\!}$ the spherical dyad introduced above \eqref{s8}. Explicitly, this background spin connection has components $\overline{\omega}_{\ha\hr}=\zeta_{\ha a}\dd\th^a$ and $\overline{\omega}_{\ha\hb}=\E{a}{a}D_b\zeta_{\hb a}\dd\th^b$. The observer thus uses flatness at infinity to adjust their frame by a rotation that is purely determined by fixed asymptotic structures $(q_{ab},\E{a}{b})$, regardless of the presence of bulk radiation.

The choice of path defining $R(\th)$ generally affects its value, but it is ultimately irrelevant: for our purposes, it suffices that $R(\th_0)=\mathbb{I}$ be the identity at the observer's initial angular position $\th_0$, \ie that the source-oriented and star-oriented tetrads initially coincide. As a result, the only relevant contribution of $R(\th)$ to the transformation law of the spin connection $\omega\to R\omega R^{-1}+R\dd R^{-1}$ stems from the inhomogeneous term, which cancels as desired the uninteresting rotation due to the observer's motion on a celestial sphere. A star-oriented tetrad has thus been defined, and one may finally use it to measure the precession of gyroscopes.

\section{Precession and memory}
\label{se3}

This section presents our results: a formula for the gyroscopic precession rate in terms of shear and news tensors, and the resulting expression for net orientation memory (see eqs.\ \eqref{t12} and \eqref{s16} below). We use the star-oriented frame $\bff{\mu}$ and focus on freely falling observers (accelerated observers are addressed in \cite{Seraj:2021rxd}).

\subsection{Precession as dual mass}
\label{se3a}

Any small freely falling gyroscope has a spin vector $\bS$ that is parallel-transported along its world line: $\nabla_{\bu}\bS=0$ in terms of proper velocity $\bu$. Now let $\{\bff{\mu}\}$ be a tetrad at the gyroscope's location such that $\bff{0}=\bu$. Then the spin vector may be written as $\bS=S^{\hi}\bff{i}$ and parallel transport becomes equivalent to a precession equation
\beq
\label{t11}
\frac{\dd S^{\hat{i}}(\tau)}{\dd\tau}
=
\Omega^{\hat{i}}_{\;\hat{j}}(\tau)\,
S^{\hat{j}}(\tau),
\ee
where $\tau$ is the observer's proper time and the angular velocity (precession rate) matrix $\Omega^{\hat{i}}_{\;\hat{j}}(\tau)$ is the projection along $\bu$ of the spin connection $\omega$ of the tetrad $\{\bff{\mu}\}$:
\beq
\label{s11}
\Omega^{\hat{i}\hat{j}}
=
-u^{\alpha}\omega_\alpha{}^{\hat i \hat j},
\quad
\omega_{\mu}{}^{\hat\mu \hat\nu}
\equiv
f^{\hat\mu}{}_{\alpha}\nabla_{\mu}f^{\hat\nu\alpha}.
\ee
The gyroscope's precession rate is thus wholly determined by the spin connection evaluated along the observer's trajectory. In practice, it is simpler to compute the spin connection of the source-oriented tetrad $\{\be{\mu}\}$ defined in eqs.\ \eqref{s7}--\eqref{s8}, then use the rotation \eqref{s9} to obtain the spin connection of $\bff{\mu}$. At leading order in $1/r$ along the world line of the observer, this yields
\begin{subequations}
\label{s12}
\begin{align}
\omega_{\hr\ha}
&\sim
\E{a}{a}\big(\tfrac{1}{4r^2} N_{ab}D_cC^{bc}\dd u
{+}\tfrac{D^bC_{ab}}{2r^2}\dd r
{+}\tfrac{N_{ab}}{2}\dd\th^b\big),\\
\omega_{\ha\hb}
&\sim
-\tfrac{1}{2r^2}\E{a}{a}\E{b}{b}\big(D_{[a}D^cC_{b]c}-\tfrac12 N_{c[a}C^c{}_{b]}\big)\dd u\nonumber\\
&\qquad+\cO(r^{-3})\dd r+\cO(r^{-1})\dd\th^a
\end{align}
\end{subequations}
where $A_{[a}B_{b]}\equiv\tfrac{1}{2}(A_aB_b-A_bB_a)$. Note that this only depends on the shear $C_{ab}$ and the news $N_{ab}$, without any influence of mass or angular momentum aspects: the latter only appear in subleading terms that are neglected here. (This notably includes Lense-Thirring precession \cite[sec.\ 40.7]{Misner:1973prb} at order $\cO(r^{-3})$.)

It is straightforward to obtain the angular velocity matrix \eqref{s11} from the geodesic velocity \eqref{s7} and the spin connection \eqref{s12}. Indeed, one finds $\Omega_{\ha\hat r}=\cO(r^{-3})$ and
\begin{empheq}[box=\othermathbox]{equation}
\label{t12}
\Omega_{\ha\hb}
\sim
\frac{\eps_{\ha\hb}}{r^2}
\tl\cM,
\quad
\tl\cM\equiv
\tfrac14 D_{a}D_b\tl C^{ab}-\tfrac18 N_{ab}\tl C^{ab}
\end{empheq}
where the {\it dual shear} tensor is $\tl C_{ab}\equiv\epsilon_{ca}C_{b}{}^c$ in terms of the Levi-Civita tensor density on the unit $S^2$. This is our first main conclusion: the precession of a gyroscope occurs, at leading order, in the plane $\ha\hb$ orthogonal to incoming radiation. Furthermore, the precession rate is proportional to a celestially local current $\tl\cM(u,\th)$, namely the {\it dual covariant mass aspect} \cite{Freidel:2021fxf,Freidel:2021qpz} closely related to the symmetry of asymptotically flat gravity under so-called dual supertranslations \cite{Godazgar:2018qpq,Godazgar:2018dvh,Kol:2019nkc,Godazgar:2019dkh,Godazgar:2020gqd,Godazgar:2020kqd,Oliveri:2020xls,Kol:2020ucd}. As announced in section \ref{sintro}, we have thus found that the precession caused by gravitational waves probes a fundamental property of the gravitational phase space. This point will be further supported in section \ref{se3b} by a relation between orientation memory and the generator of gravitational electric-magnetic duality.

For future reference, it is useful to write tensors on celestial spheres in terms of (pseudo-)scalar functions with definite parity. Let therefore angular momentum and shear be written as
\begin{subequations}
\label{s13}
\begin{align}
L_a
&\equiv
D_aL^++\eps_{ab}D^bL^-,\\
\label{ss13}
C_{ab}
&\equiv
D_{(a}D_{b)}C^{+}
-\tfrac12q_{ab}D^2C^+
+\epsilon_{c(a} D_{b)} D^{c} C^{-},
\end{align}
\end{subequations}
where $A_{(a}B_{b)}\equiv\tfrac{1}{2}(A_aB_b+A_bB_a)$ and the superscript $+$ (resp.\ $-$) denotes even (resp.\ odd) functions. The term linear in $C$ in Eq.\ \eqref{t12} can then be recast as $\tfrac18D^2(D^2+2)C^-$, which is manifestly odd. Furthermore, the balance equation \eqref{tt6} allows us to relate this term to the time derivative of the odd component of angular momentum and its flux as
\beq
\label{s14}
\tl\cM
=
\dot L^-
+\cJ^-
-\tfrac18N_{ab}\tl C^{ab},
\ee
where $\cJ^-=D^{-2}(\eps^{ab}D_b\cJ_a)$ and $D^{-2}$ is the inverse of the Laplacian on $S^2$, involving the Green's function $G$ such that $D^2G(\th,\th')=\tfrac{1}{\sqrt{q}}\delta(\th-\th')-\tfrac{1}{4\pi}$. This
confirms the expected absence of precession in non-radiative space-times, since $N_{ab}=0$ also implies $\dot L^-=\cJ^-=0$ (at least in the absence of NUT charges \cite{Kol:2020ucd}). Relative to fixed distant stars, precession thus occurs only during the passage of a wave. It is therefore meaningful to compute the net change of orientation due to a burst of radiation.

\subsection{Orientation memory}
\label{se3b}

One can readily write the solution of the precession equation \eqref{t11} as a time-ordered exponential of the matrix $\Omega$ acting on some initial spin $\bS_{\text{init}}$. In practice, only the first non-trivial term of this expansion is reliable, since the angular velocity \eqref{t12} is of order $\cO(1/r^2)$ anyway and higher-order terms of the exponential series are affected by subleading $1/r$ corrections that have been neglected here. Accordingly, the leading-order change of orientation of the gyroscope's axis is given by $\Delta S^{\hr}=\cO(r^{-3})$ and $\Delta S^{\ha}=\Phi\eps^{\ha\hb}S^{\hb}_{\text{init}}+\cO(r^{-3})$, where the rotation angle in the $\ha\hb$ plane is obtained by integrating the covariant dual mass aspect \eqref{s14} over time:
\beq
\label{s15}
\Phi
=
\int\!\dd u\,\frac{\tl\cM}{r^2}
=
\tfrac{1}{r^2}\Big[\Delta L^-+\int\!\dd u\big(\cJ^--\tfrac{1}{8}N_{ab}\tl C^{ab}\big)\Big].
\ee
The fact that $\Phi\neq0$ is the aforementioned memory effect: the passage of waves entails a permanent change of orientation, sensitive to a specific combination $\tl\cM$ of metric components. The latter can be rewritten more suggestively by `inverting' the parity decomposition \eqref{s13} under the assumption (without loss of generality) that $L^{\pm}$ have vanishing average on $S^2$ while $C^{\pm}$ both have vanishing harmonics of order $\ell=0,1$. This rephrases the memory effect \eqref{s15} as
\begin{empheq}[box=\othermathbox]{equation}
\label{s16}
\Phi
=
\frac{8\pi}{r^2}
\Big(\Delta Q_Y+\cF_Y-\tfrac{1}{64\pi}\int\dd u\,N_{ab}\tl C^{ab}\Big),
\end{empheq}
where all terms on the right-hand side are evaluated at $\th$ and we have introduced a divergence-free vector field $Y^a(\th')\equiv\eps^{ab}D_bG(\th,\th')$, while $Q_Y\equiv\tfrac{1}{8\pi}\oint\sqrt{q}\dd^2\th'\,Y^aL_a$ is the associated super-angular momentum charge \cite{Compere:2019gft} and $\cF_Y\equiv\tfrac{1}{8\pi}\oint\sqrt{q}\dd^2\th'\,Y^a\cJ_a$ is its flux. This makes it manifest that the first two terms of gyroscopic memory (namely $\Delta Q+\cF$) reproduce the spin memory effect \cite{Pasterski:2015tva}. Crucially, however, Eq.\ \eqref{s16} involves an additional nonlinear term $\propto\int N_{ab}\tl C^{ab}$; the latter is nothing but the Hamiltonian generator of local electric-magnetic duality transformations on radiative phase space endowed with its standard symplectic form \cite{Ashtekar:1981bq}. This exhibits once more the deep relation between gyroscopic memory and crucial gravitational symmetries.

The non-vanishing value of \eqref{s16} also illustrates the general memory mechanism suggested in Fig.\ \ref{fiMem}. In the context of displacement memory \cite{Strominger:2014pwa}, the `metric perturbation' of Fig.\ \ref{fiMem} is the shear $C_{ab}$ and the net change $\Delta C_{ab}$ measures the angular deviation of nearby geodesics. Gyroscopic memory is more subtle in that respect, as the perturbation should now be seen as a time integral of dual shear through the dual covariant mass aspect of Eq.\ \eqref{t12}. The presence of such time-integrated metric perturbations is typical of subleading memory effects \cite{Nichols:2017rqr,Nichols:2018qac,Seraj:2021qja}.

To conclude, let us estimate the magnitude of the memory effect \eqref{s16}: it falls of as $1/r^2$ and is in this sense dominant with respect to Lense-Thirring precession, which falls off as $1/r^3$. Could it then be possible to observe the precession described here? The answer is unclear at the moment, as realistic values of \eqref{s16} are tiny. Indeed, elementary dimensional analysis suggests that the order of magnitude of \eqref{s16} for a bound binary system with mass scale $M$ is
\beq
\Phi
\sim
\frac{G^2}{c^4}\frac{M^2}{r^2}
\simeq
2\times 10^{-39}\Big(\frac{M/M_{\odot}}{r/1\,\text{Mpc}}\Big)^2
\ee
where $G$ is Newton's constant, $c$ is the speed of light in vacuum and $M_{\odot}$ is the solar mass. This is manifestly exceedingly weak for the common values of mass and distance ($M\simeq30M_{\odot}$ and $r\simeq400\,\text{Mpc}$ for the seminal LIGO observation \cite{LIGOScientific:2016aoc}). It is nevertheless conceivable that the effect will some day be observable in extreme events such as supermassive black hole mergers, for which values of the order of $\Phi\simeq10^{-26}\,\text{rad}$ are not far-fetched. Also note that the effect is independent of the gyroscope's spin and inertia, so one may even resort to distant pulsars (whose high mass and low volume reduce non-gravitational environmental effects) as radiation probes. Small rotations of a pulsar's axis could then conceivably be measured: for instance, a $\Phi\sim 10^{-26}\,\text{rad}$ change in the angle of an idealized narrow beam emitted $10^3$ light-years away from the Solar system modifies the position of the resulting light spot on Earth by about $10^{-7} m$. We hope to further develop some of these ideas in future work: it would be fascinating indeed to observe gravitational memory effects with the simple gyroscopic setup described here.

\section*{Acknowledgements}

We are grateful to Glenn Barnich, Fran\c{c}ois Mernier and Roberto Oliveri for insightful comments on an early draft of this \letter, and to Miguel Paulos for suggesting to use pulsars as gyroscopes. The work of A.S.\ is funded by the European Union’s Horizon 2020 research and innovation programme under the Marie Sk{\l}odowska-Curie grant agreement No.\ 801505. The work of B.O.\ is supported by the ANR grant {\it TopO} No.\ ANR-17-CE30-0013-01, and by the European Union's Horizon 2020 research and innovation programme under the Marie Sk{\l}odowska-Curie grant agreement No.\ 846244.

\addcontentsline{toc}{section}{References}
\providecommand{\href}[2]{#2}\begingroup

\end{document}